\documentclass[3p,times]{elsarticle}

\usepackage{ecrc}

\volume{00}

\firstpage{1}

\journalname{Procedia Computer Science}

\runauth{}

\jid{procs}

\jnltitlelogo{Procedia Computer Science}

\CopyrightLine{2011}{Published by Elsevier Ltd.}

\usepackage{amssymb}

\usepackage[figuresright]{rotating}

\usepackage[utf8]{inputenc}
\usepackage[T1]{fontenc}
\usepackage{framed}
\usepackage{multirow}
\usepackage{color}
\usepackage{graphicx}
\usepackage{tabulary}
\usepackage{enumitem}
\usepackage{hyperref}
\usepackage{url}
\usepackage{soul}
\usepackage{rotating}
\usepackage{tcolorbox}
\usepackage{lscape}
\usepackage{caption}
\usepackage[title]{appendix}
\usepackage{ulem}
\usepackage{placeins}
\usepackage{balance}
\usepackage[autostyle, english = american]{csquotes}
\MakeOuterQuote{"}

\begin{document}

\begin{frontmatter}



\dochead{}

\title{A City upon a Hill: Casting Light on a \textit{Real} Experimental Process}


\author{Efraín R. Fonseca C.}
\address{Av. General Rumiñahui s/n y Ambato Sangolquí – Ecuador}

\author{Marta López-Fernández}
\address{Calle Profesor José García Santesmases 9, 28040, Madrid, España}

\author{Oscar Dieste}
\address{Calle de los Ciruelos, 28660 Boadilla del Monte, Madrid, España}

\author{Natalia Juristo}
\address{Calle de los Ciruelos, 28660 Boadilla del Monte, Madrid, España}


\begin{abstract}
\textbf{Context:} The overall scientific community is proposing measures to improve the reproducibility and replicability of experiments. Reproducibility is relatively easy to achieve. However, replicability is considerably more complex in both the sciences and Empirical Software Engineering (ESE). Several strategies, e.g., replication packages and families of experiments, have been proposed to improve replication in ESE, with limited success. We wonder whether the failures are due to some mismatch, i.e., the researchers' needs are not satisfied by the proposed replication procedures.\\
\textbf{Objectives:} Find out how experimental researchers conduct \textit{experiments in practice}.\\
\textbf{Methods:} We carried out an ethnography study within a SE Research Group. Our main activity was to observe/approach the experimental researchers in their day-to-day settings for two years. Their preferred literature and experimental materials were studied. We used individual and group interviews to gain understanding and examine unclear topics in-depth.\\
\textbf{Results:} We have created conceptual and process models that represent how experimentation is \textit{really} conducted in the Research Group. Models fit the community's procedures and terminology at a high level, but they become particular in their minute details.\\
\textbf{Conclusion:} The actual experimental process differs from textbooks in several points, namely: (1) Number and diversity of activities, (2) existence of different roles, (3) the granularity of the concepts used by the roles, and (4) the viewpoints that different sub-areas or families of experiments have about the overall process.
\end{abstract}


\end{frontmatter}



\section{Introduction}\label{sec-introduction}
Empirical Software Engineering (ESE) methods and procedures develop over time, e.g., \cite{Kitchenham-2016-robust, Arcuri-2014-randomized, Ferreira-2017-planning-experiments, Fonseca-2017-design-experiments, Anchundia-2020-resources-reproducibility}. However, improvements are incremental. For instance, replication is one ESE area that invests much effort. However, the number of replications is small \cite{Bezerra-2015-Replication-SE-U-SMS}, and the rate of confirmation of previous results is limited \cite{Jorgensen-2016-Incorrects-Results-SEE}. Such a situation is not privative of ESE. Moreover, replication is troublesome in general \cite{Klein-2018-many}, particularly in the social and life sciences \cite{Pashler-2012-perspectives, Baker-2016-lid-reproducibility}.

Many potential reasons prevent effective experimental replication. They have discussed elsewhere, e.g., \cite{Miller-2005-replicating-SE-experiments, Demagalhaes-2015-replications-SE}. In this paper, we focus on a potential cause that, to the best of our knowledge, has not been adequately explored: How ESE researchers \textit{really} experiment.

Anecdotic evidence such as conversations with other researchers, review of articles, direct observation, among others, suggests that researchers follow \textit{in practice} different \textit{experimental processes}, i.e., the different activities that ESE researchers perform to conduct experiments. The experimental reports look relatively uniform due to the existence of reporting guidelines \cite{Carver-2010-guidelines-replication-SE, Jedlitschka-2008-reporting-experiments-SE}, but the underlying data-generation processes may be somewhat different.

This paper aims to find out \textbf{how experimental researchers conduct experiments in practice}. We applied an ethnographic method \cite{Sharp-2016-Ethnographic-Studies-ESE} within an experienced ESE research group. We observed/approached the experimental researchers in their day-to-day settings for two years. Their preferred literature and experimental materials were studied. We used individual and group interviews to gain understanding and examine unclear topics. 

We created conceptual and process models to represent the experimentation's process in a \textit{concrete} Research Group. Models fit the community's procedures and terminology at a high level. However, they become particular in their minute details: (1) Number and diversity of activities, (2) existence of different roles, (3) the granularity of the concepts used by the roles, and (4) the viewpoints that different sub-areas or families of experiments have about the overall process. The contributions of this paper are:

\begin{itemize}
    \item Conceptual and process models that represent how experimentation is conducted \textit{in practice} in an ESE research group.
    \item The identification of roles and activities that \textit{significantly} depart from the usual experimentation standards. 
    \item Opening a discussion in the ESE community: Should those differential characteristics be disregarded as inherently contextual? Alternatively: Are they more common than it seems, so they should be further investigated and eventually integrated into mainstream practice?
\end{itemize}

The remainder of the article is structured as follows: Section \ref{sec-research-method} describes the research method followed. Section \ref{sec-reseach-execution} characterize in detail the investigation developed in this work. Section \ref{sec-related} discusses the models structured and possible future actions. Section \ref{sec-threats} details threats to validity and its prevention. Finally, in section \ref{sec-conclusions}, the conclusions around the research are presented.
\section{Research Method}\label{sec-research-method}
We selected Ethnography as the research method due to it is characterized by the \textit{participant observation} method, compared to other qualitative research methods. In addition, ethnographic researchers adopt \textquotedblleft the native's point of view\textquotedblright~to better understand the reason for their actions \cite[pp. 55-56]{Outhwaite-2007-sage}. Thus, the biases and misperceptions that ESE researchers might have about themselves can be detected. Ethnography also captures the rich context that surrounds ESE researchers.

Given the high cost and effort that ethnographical research represents, we selected a representative ESE research group to guarantee, to the extent possible, the reliability of the results obtained, their validity, and scientific significance \cite{sjoberg-2007-future-empical-methods}. We followed the Runenson et al.'s case studies guidelines \cite{Runenson-2012-case-study-SE} wherever they applied to ethnographical studies through this protocol:

\begin{itemize}
	\item{Context selection, that is, to select the context that provides data and facilities for the research in the best possible condition.}
    \item{Data collection procedures, that is, to specify the process, tools, and materials focused on collecting data.}
    \item{Data analysis procedures, that is, particularize the techniques, tools, and activities selected for analyzing data.}
\end{itemize}

To avoid confusion, we will refer to the~\textquotedblleft ESE researchers\textquotedblright~with whom we interacted with during the investigation as \textquotedblleft the experimenters\textquotedblright. We will talk about themselves the authors as the \textquotedblleft the researchers\textquotedblright. 

\subsection{Context selection}
We have considered an ESE research group as ''representative'', i.e., a reliable source of information, when it fulfills the following context selection criteria:

\begin{itemize}
	\item{The research group has extensive experience in ESE research.}
	\item{The ESE community recognized the experimenters as specialists.}
	\item{The group has a large number of publications in journals of repute and internal documentation, i.e., the group has a verifiable track record.}
\end{itemize}

\subsection{Data collection procedures}\label{subsec-data-collection}
The information sources within the ESE research group under study (RGUS) are:

\begin{itemize}
	\item Experimenters' knowledge.
	\item General and specific literature\footnote{E.g., the book \textquotedblleft Experimentation in Software Engineering\textquotedblright~\cite{Wohlin-2012-experimentatio-SE}.}. 
	\item General and specific experimental material, e.g., experimental designs.
	\item Daily or exceptional activities carried out, e.g., planning of experiment activities).
\end{itemize}

The order in which researchers contact information sources is random. The frequency of the contact depends on the researchers' needs to obtain knowledge to answer the questions that show up during the research, and also on the availability of each source. 

Ethnography information gathering is characterized by being cyclical, iterative, and incremental. The data acquisition process was based on a taxonomy of information gathering techniques, categorized according to the level of contact with the primary source of information. Such taxonomy proposes three levels of information gathering: direct participation with the source (level 1), indirect participation with the source (level 2), and study of the experiment material without the participation of the source (level 3) \cite{Lethbridge-2005-data-collection-techniques-SE}. Table \ref{tbl-tecnica-fuente} describes the information-gathering techniques and instruments, and the corresponding information sources to which they can be applied to.

\begin{table}
	\small
	\centering
	\caption{Techniques and tools of collection by source of information}
	\label{tbl-tecnica-fuente}
	\begin{tabular}{|p{2cm}|p{3.5cm}|p{5cm}|}
	\hline
	\textbf{Source} & \textbf{Technique} & \textbf{Instruments}\\
	\hline
	\multirow{7}{50 pt}{Experimenters knowledge} & Interviews & Video camera, paper, pencil\\
	\cline{2-2}\cline{3-3}
	& Focus groups & Video camera, blackboard, cardboard, board markers, chalkboard eraser, etc.\\
	\cline{2-2}\cline{3-3}
	& Process modeling & Video camera, paper, pencil\\
	\cline{2-2}\cline{3-3}
	& Instrumental systems & Email, videoconferences, internet\\
	\cline{2-2}\cline{3-3}
	& Experience-based learning & Literature review, review of experimental material, experimentation in practice\\
	\hline
	\multirow{3}{15 pt}{Common literature} & Documentation analysis & Comprehensive reading of literature, reading summaries, software support tools\\
	\hline
	\multirow{2}{15 pt}{Experimental material} & Documentation analysis & Systematic analysis of experimental material, trials with data\\
	\hline
	\multirow{2}{15 pt}{Group activities} & Participatory observation & Literature review, researchers knowledge\\
	\hline
	\end{tabular}
\end{table}

\subsection{Data analysis procedure}

The data analysis procedure consists of four activities: 
\begin{itemize}
\item Preparation of materials: The information sources are classified and categorized in such a way that both experimenters and researchers can manage it.

\item Information acquisition: During the analysis of the information sources, researchers gradually acquire basic knowledge about experimentation, which helps them to more easily understand the experimenter's knowledge and behavior.

\item Verification of information: This is probably the most complex activity. The knowledge obtained from the sources complements each other. Triangulation will be used to explain the findings and identify new areas of inquiry. All findings are discussed with the experimenters to avoid misunderstandings and verify that the resulting activities are those that they really carry out in practice.

\item Generation of results: Finally, the results are obtained as final products after several iterations of comparison, verification, and particularly consensus between researchers and experimenters. The expected results will be models (conceptual and process) and the practices derived from the interviews with the experimenters.
\end{itemize}
\section{Ethnographic study}\label{sec-reseach-execution}
We will report the study's results in chronological order: First, we describe the results of the RGUS's preferred sources to review. Afterward, the product of a specific experimental material review is detailed. Finally, we explain the outcomes of formalizing the experiment terminology in SE and interviews with the RGUS's researchers. All data collected during the fieldwork (photographs, audio recordings, videos, and reports) are available under request.

\subsection{First Approximation}
Researchers started their activities by conducting interviews with experimenters of high and low experience. These activities aimed to make explicit the RGUS' ESE knowledge and convert it into models that allow analysis and subsequent manipulation.

Interviews and workshops were not enough to achieve the research objectives. The RGUS members provided only explicit knowledge about an experiment's general components. They believed that the interviews were unnecessary because the researchers could acquire the knowledge sought in the literature that experimenters considered the standard, e.g., \cite{Wohlin-2000-Experimentation-SE}.

When the research progressed into specialized areas, communication became difficult. The experimenters supported their story-telling in specific experimental materials that the researchers did not have. Nor did the researchers have the level of knowledge required to understand the experimenters' discussions. Consequently, the researchers carried out additional activities to understand the standard literature and experimental material, while the interaction with the experimenters continued at a slower pace. For example, the researchers were experimental subjects and then knowledge managers of an RGUS's family of experiments.

\subsubsection{Analysis of the relevant literature}
The experimenters use the books by Wohlin et al. \cite{Wohlin-2000-Experimentation-SE}, and Juristo \& Moreno \cite{Juristo-2001-SE-experimentation} as primary sources for their experimentation process; as well as specific papers that report how to experiment in SE, e.g., Basili et al. \cite{Basili-1986-ESE}, and Kitchenham et al. \cite{Kitchenham-2002-empirical-research-guidelines-SE}. The work of Basili et al. \cite{Basili-1999-families-experiments} was also considered relevant due to the experimenters' interest in carrying out experimental replications.

The researchers conducted an in-depth literature review, which allowed them to identify a marked terminological diversity. The different sources refer to similar activities of the SE experimentation process. In the beginning, it seemed to the researchers that the literature describes other experimental procedures. Time later, they verified that the different terms refer to essentially the same activities. By way of example, the materials cited above designate the phases of the SE experimentation process as follows:

\begin{itemize}
	\item Wohlin et al. \cite{Wohlin-2000-Experimentation-SE}: \textquotedblleft Experiment idea, Experiment scoping, Experiment planning, Experiment operation, Analysis \& interpretation, Presentation \& package, and Experiment report\textquotedblright.
	\item Juristo \& Moreno \cite{Juristo-2001-SE-experimentation}: \textquotedblleft Objective Definition, Design, Execution and Analysis\textquotedblright.
	\item Basili et al. \cite{Basili-1986-ESE}: \textquotedblleft Definition, Planning, Operation, and Interpretation\textquotedblright.
	\item Kitchenham et al. \cite{Kitchenham-2002-empirical-research-guidelines-SE}: \textquotedblleft Experimental context, Experimental design, Conduct of the experiment and Data collection, Analysis, Presentation of results, and Interpretation of results\textquotedblright.
\end{itemize}

The researchers looked for additional ESE books to verify whether terminological diversity was a widespread problem. However, we only found the new edition of the book by Wohlin et al. \cite {Wohlin-2012-experimentatio-SE} on experimentation, and the specialized books of Runeson et al. \cite{Runenson-2012-case-study-SE}, and Kitchenham et al. \cite{Kitchenham-2015-Evidence-Based-SE}. The ESE-specific literature is thus somewhat limited.

\subsubsection{Review of experimental material}\label{subsubsec-review-experimental-material}
The RGUS' experimental materials were a promising source of information since they were generated during almost two decades of experimentation and reflect the tacit knowledge \cite{Polanyi-1996-tacit-k} present in the group. However, its access was challenging due to shows disorganization. We needed the experimenters' help to find specific materials, understand their structure and contents, and gain further access to other related materials.

The experimental materials evolved. Experiments vary widely in documentary support, independently of their date. The experimental family explains the differences better than the time; the materials related to different families of experiments show significant differences, e.g., digital raw data versus physical papers written with raw data; or automatic tools to measure variables versus manual tools.

Concerning the experimental process, all experimental materials describe the activities that have to be carried out in the experiment in a general way. The order of execution and the relationship between experimental activities are not clearly defined; they rely on the experimenters' tacit knowledge. Experimental materials are also affected by terminological diversity, especially among families of experiments.

We intended to acquire the RGUS' knowledge and drain it into process and conceptual models. However, we largely failed to consolidate the terminology, so the resultant model, e.g., see this 
\href{https://drive.google.com/open?id=1EzfWnEy-_9evvuGum_xEjG6CqVBI2cQt}{\ul{link}},
was relatively simple. Even so, this abstract preliminary model was adequate to continue with the ethnographic research.

\subsection{Diving into the SE experimental process}\label{replication}
The theoretical concepts acquired by the researchers were inconclusive without some practical confirmatory application. We decided to learn-by-doing conducting an experiment ourselves.

\subsubsection{Experiment selection}
The researchers opted to replicate an experiment previously conducted by the RGUS, instead of designing an experiment from scratch. The replication attracted the attention of experimenters and guaranteed their collaboration. Furthermore, the replication would give us the chance to recreate the design, execution, analysis, and reporting of an experiment that, until then, we only knew due to the examination of the written material. Finally, the replication would help us to better understand the experimenters' motivations and decisions.

\subsubsection{Replication planning}
Diverse problems hinder replication \cite{Gallardo-2012-CG-PL-SE, Vegas-2006-communication-researchers, Miller-2005-replicating-SE-experiments, Gomez-2014-understanding-replication, Demagalhaes-2015-replications-SE, Carver-2010-guidelines-replication-SE}. In our case, as mentioned before, the main problem was the lack of a detailed account of the experimental activities, e.g., how to approach the experimental design, which experimental objects were necessary and in which order, etc. We requested the cooperation of the experimenters (who ran the baseline experiment) to schedule the experimental activities and adapt the design to the new context where the replication would be run.

\subsubsection{Conduction}
One researcher performed the replication in an external academic setting to RGUS. The existence of explicit experimental materials such as questionnaires, task descriptions, programs, among others, significantly eased experiment conduction since the researchers defined in advance all relevant aspects.

The measurement was the most complex and controversial aspect. The experimental subjects generate raw data using test cases previously proposed, but such test cases are manually interpreted, not exercised. The cooperation of the experimenters was again needed to complete the measurement. Even if the measurement procedure looked controversial, it made complete sense after discussion. We learned that the experimental protocols strongly depended on the experimental task, which depends on the research area. The remaining steps, e.g., calculation of measurements from raw data,  formatting, etc., did not imply any uncertainty. It was possible to reproduce the same files used in the baseline experiment.

\subsubsection{Analysis and reporting}
We sought help to analyze and report the replication since we did not have statistical training. In retrospect, both the analysis and the report were simple to perform since the experimental design used (\textit{cross-over}) was associated with a well-defined analysis method \cite{Vegas-2016-crossover-designs-experiments-SE}. On the other hand, there were reporting guidelines available \cite{Carver-2010-guidelines-replication-SE}. The experimenters confirmed that generally, the analysis and reporting of experiments do not imply substantial SE challenges.

The differences observed between replication and the baseline experiment confirmed our belief that, at least in the case of the RGUS, there are no clear guidelines that allow literal, external, and independent replication of the original experiment, irrespective of the existence of incidents in the replication's execution. As a corollary, original experimenters' presence is necessary to make the baseline experiment results and the replication comparable.

\subsubsection{Assessment}
The replication was, in retrospect, the key that clarified the experimental process. The experience increased the researchers' knowledge, which enabled the verification of previous findings, e.g., the little detail of the experimental materials). New findings were also made, e.g., the specificity of experimental protocols and, notably, the confirmation of extensive tacit knowledge.

\subsection{Mining the tacit knowledge}\label{subsec-conocimiento-grupo}
After the experiment, the researchers were not sure which way to go. The investigation was a success from an operational viewpoint, except the analysis, which required specific statistical knowledge. The researchers satisfactorily completed all other activities. We increased the confidence in our experimental ability. The models did not improve much during the replication, e.g., see this \href{https://drive.google.com/open?id=1CZKiYhO6hqq29IwSB-7m_HBO5JlgQtWE}{\ul{link}}, but it seemed that we did not overlook anything relevant.

However, the presence of tacit knowledge was a research's threat to validity. We were able to replicate \textbf{one} experiment in \textbf{one} concrete SE area, but nothing guaranteed that we would not need the experimenters' help in the subsequent replication. According to the SE literature \cite{Juristo-2012-replication-SE,Gomez-2014-understanding-replication}, it looked just the opposite.

We decided to organize new series of semi-structured interviews with the experimenters. The key difference with the first approximation was that we would not discuss the SE experimental process \textit{in general}, but we would focus on \textit{specific families of experiments}. Given the events during the experiment, we believed that this strategy could expose specific experimenters' knowledge.

\subsubsection{Semi-structured interviews}
\begin{figure}[htbp!]
	\centering
	\includegraphics[width=\columnwidth]{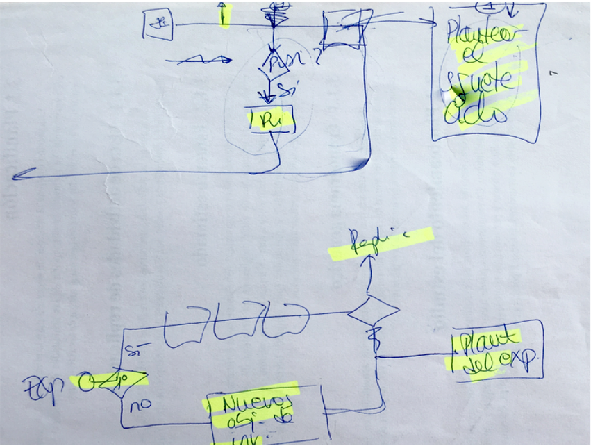}
	\caption{Example of a sketch made by experimenters.}
	\label{fig-proceso-exp-boseto}
\end{figure}

We created a questionnaire based on open-ended questions to motivate spontaneous in the experimenter, but relatively ordered narrative about his/her experiences. We abandoned note-taking and went on to capture on video the details of the experimenter's verbal and gestural expression
. We also collected the paper sketches (e.g., see Fig.~\ref{fig-proceso-exp-boseto}) that some experimenters used to support their narrative. We subsequently analyzed all this material following a process similar to protocol analysis \cite{Pressley-1995-verbal-protocols}. The researchers created a report after each interview, e.g., see this \href{https://drive.google.com/file/d/1jc4KOrDMtxvqSNaIjRRE_ycTP7gByW7z/view?usp=sharing}{\ul{link}}.

\begin{figure}[htbp!]
	\centering
	\includegraphics[width=\columnwidth]{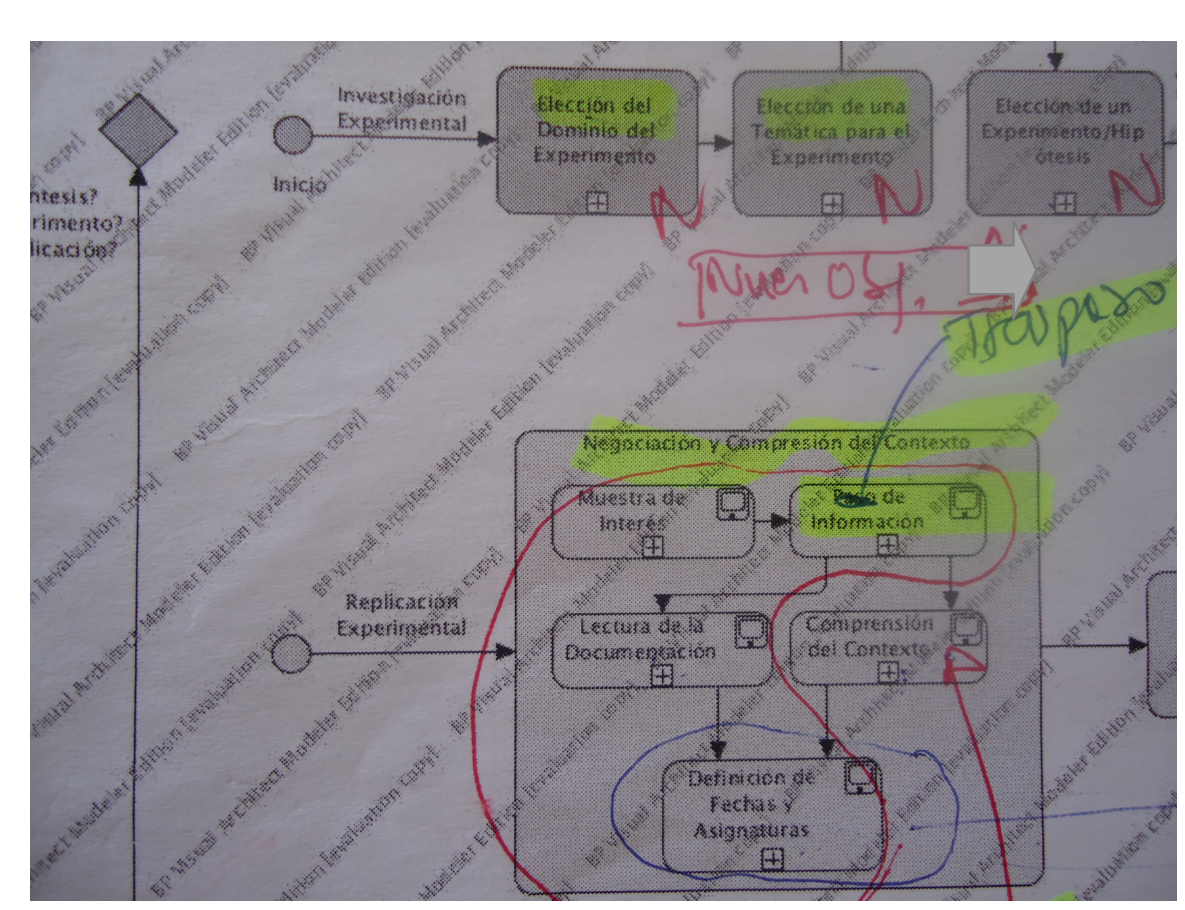}
	\caption{Extract from the experimental activities carried out by the RGUS experimenters (intermediate version). The evolution regarding Fig.~\ref{fig-proceso-exp-boseto} is observable in terms of the number of activities performed and transparency in the workflow.}
	\label{fig-proceso-draft}
\end{figure}

The experimenters validated the knowledge obtained in each session at the beginning of a subsequent session. We conducted a total of seven interviews with six RGUS members. The process was smooth since each experimenter offered a coherent narrative. In this way, different iterations allowed to progressively refine the acquired knowledge up to a point where progress was marginal, e.g., see Fig.~\ref{fig-proceso-draft}. At that moment, the interviews concluded. The reader can check the models' evolution through this \href{https://drive.google.com/drive/folders/1QgUGzp5RGQ_BzV_ggvCgZRl1Znj5o28Q?usp=sharing}{\ul{link}}.

The researchers summarized the acquired knowledge in three intermediate research products:
\begin{itemize}
\item \textit{Preliminary} experimental process model: We realized that the SE experimental process had the same characteristics as any other process, in particular, the SE processes (e.g., ISO/IEC 12207 \cite{ISO-IEC-IEEE-12207}). Hence, we identified six high-level experimental processes: support, generation of pieces of knowledge, publications, synthesis, basic and organizational, which could see in this \href{https://drive.google.com/file/d/1NfemxVrp-D_ss8HOJ9L_RAEVGIF185c5/view?usp=sharing}{\ul{link}}.
\item \textit{Preliminary} workflow: It shows the sequence of activities to be carried out in each of the main experimentation processes. See details in this \href{https://drive.google.com/file/d/1zm8JdDIooUa2IcAiKEagCU3DUFG_WecP/view?usp=sharing}{\ul{link}}.
\item \textit{Preliminary} conceptual model: The interviews do not generate new conceptual knowledge, i.e., the knowledge acquired during the replication was essential.
\end{itemize}

The interviews showed that a single experimenter does not usually carry out all the SE experimental process's activities. Responsibilities are assigned to different roles within a research group. Roles may be played by the same experimenter or, as in the RGUS, by different experimenters. We identified the following roles (names suggest their responsibilities): (1) \textit{Research Manager}, (2) \textit{Experiment Manager} and (3) \textit{Senior Experimenter}. These roles are responsible for: (1) planning the experimental research, (2) conducting the experiment in practice, and (3) managing the logistics, respectively. The study showed some overlapping of responsibilities among roles;  such overlapping has surely an impact (positive or negative, depending on the case) in experimental management.

\subsubsection{Joint focus groups}\label{subsubsec-focus-groups}
When the preliminary process and models were ready, we made a presentation to all RGUS' experimenters. The researchers organized meetings around families of experiments up to this point, which was an unfortunate mistake. The experimenters agreed again at the general level, but they disagreed in the details, typically along the experimental families' conceptual borders. For instance, the experimenters never reached an agreement on activities around the experimental design.

We were pretty sure that the interview exercise would be enough for model construction. Interviews are not that good at eliciting tacit knowledge; that is why we established extra measures such as recording videos, creating interview reports, and validate those reports with the experimenters.

The final step was to solve the disagreements using some consensus-reaching technique. We could use DELPHI \cite{Dalkey-1967-Delphi}, but social pressure did not seem a problem; all experimenters defended their positions gracefully. We decided to hold focus groups to reach a consensus or, at least, reveal the origin of the discrepancies. The most senior experimenter led the meetings as a moderator, but only to keep the discussion on track. The researchers only observed.

Each meeting lasted at least two hours. Before the focus groups, the experimenters reviewed the models, not the process, which was later updated based on the models' changes. The focus groups prompted a considerable increase in the models' details, after heated discussions that re-confirmed, if necessary, the existence of terminological diversity \textit{even among experimenters who carry out similar activities within the same family of experiments}. 

We learned during the process that the reason for the terminological diversity among experimenters seems to be the training that each experimenter received. Talking about~\textquotedblleft training\textquotedblright~in the context of the RGUS is probably incorrect since its members are self-educated, with few exceptions. SE materials did not drive self-education, few existed when the experimenters got their education, and few exist nowadays. The researchers chose the materials according to each experimenter's field of specialization. Such field was somehow related to the experimental families in which the experimenter participated. Therefore, terminological diversity was an inevitable consequence.

Finally, the experimenters reached an apparent consensus, e.g., see the model in this \href{https://drive.google.com/file/d/1rGhYwH5I99ovmFGlX4FcLDCvd-y-ajp_/view?usp=sharing}{\ul{link}}). The experimenters do not fully achieve consensus in the focus groups; we needed to carry out meetings in parallel to ''put out'' the argumentative flames. In any case, some experimenters still disagreed about the consensus model because it did not represent specific concepts that they used in their research activities.

\subsubsection{Role-focused workshops}\label{subsubsec-focus-groups-role}
After the joint focus groups, the researchers were convinced that the absolute consensus was impossible. The discussions with experimenters of different families led to a characteristic circular pattern: An experimenter triggered a change that, later, was undone by another. Substantial agreement was only possible \textit{within each family of experiments}. Not in vain, the progress that we achieved during the replication (see section \ref{replication}) was possible due to its membership to a concrete family and the support of that family's experimenters.

We did not aim to explore SE family-specific experimental processes. Albeit enjoyable, we perceived it as future research. In turn, we tried to take advantage of the available time, limited at that stage of the research, to improve the models switching the workshops' perspective to the \textit{experimental roles}.

experimenters who played the same role and analyzed such role \textit{only}. Meetings were peaceful and conversations constructive. Putting aside family-specific issues, we were able to re-validate the existence of roles (the experimenters themselves agreed that roles made sense, even if they have never perceived their responsibilities in these terms). We were also able to break down the conceptual model into role-specific conceptual models available through these  links: \href{https://drive.google.com/file/d/1RRDbJaav0XD7Oq9oG77VNTttvIZm89NL/view?usp=sharing}{\ul{1}}, \href{https://drive.google.com/file/d/1Du5viBUKfrNehlLs9U0WeknKAQzstxHe/view?usp=sharing}{\ul{2}}, \href{https://drive.google.com/file/d/1n3CHZbyjNovlgIhyTdCtN51f2X9Z-lwx/view?usp=sharing}{\ul{3}}. It is noticeable the level of detail of such models that we will describe in the next section.
\section{Results}\label{sec-results}

The research project yielded three main results: (1) an \textit{experimental process workflow}, (2) an \textit{experimental research concept model}, and to a lesser extent, (3) an \textit{experimental process model}. These models\footnote{We have not used standard notations, e.g., UML or BPMN, because experimental researchers have different ages, backgrounds, and careers. We have valued communication above model formality.} are described below.

\subsection{Experimental process workflow}

The so-called \textit{experimental process workflow} (EPW) represents the tasks carried out during SE experimentation (according to the RGUS' perspective) and their logical sequence. The workflow is available in this \href{https://drive.google.com/file/d/18-036nC2hWNBprns7Be6TBjxNK5fY2OW/view?usp=sharing}{\ul{link}} and, with lower resolution, it is shown in Fig.~\ref{fig-EPM-construction}.

The EPW model shows three well-defined paths corresponding to the key SE experimental research processes: (1) Experimentation (purple), (2) replication (green) and, (3) synthesis (yellow). It is noticeable that processes associated with planning new cycles of experimental research and publication of results (highlighted in blue) also appear associated with the experimentation and synthesis paths.

\subsubsection{Experimentation path}

Experimentation is the heart of experimental research since replication and synthesis stem from it. However, in some circumstances, experimentation may require the previous execution of synthesis (e.g., when choosing the topic of the experiment) or replication (e.g., in adaptation to the context) activities.

According to the RGUS' experimenters, the experimentation phases are:
\begin{itemize}
\item Selecting the experiment domain.
\item Selecting the experiment topic.
\item Selecting the experimental hypothesis.
\item Planning the experiment.
\item Creating the experimental kit.
\item Executing and post-execution of the experiment.
\end{itemize}

When the experiment results do not meet the experimenters' expectations, the experiment can be redesigned and conducted again. This redesign activity takes place, particularly when investigating fundamental topics.

\subsubsection{Replication path}

Replication is the practice on which scientific knowledge is based, hand in hand with the synthesis of results \cite{Roizer-2014-reproducibility}. The RGUS entirely agrees that replication is essential for experimental research both explicitly (RUGS' experimenters claim that) and, more importantly, implicitly (the experimental workflow contains minute detail about replication activities). The replication path consists of the following activities: 
\begin{itemize}
\item Definition of replication.
\item Study and update the experimental kit.
\item Adaptation to the context, and
\item Creation of operational elements.
\end{itemize}

\subsubsection{Synthesis path} 

As mentioned above, the synthesis of results is also a fundamental path in experimental research. However, our findings are just partial as the RGUS did not have any experimenter with expert knowledge available during this research.

According to the RGUS' experimenters, the synthesis of results can start from systematic literature reviews or systematic mapping studies; or the results of experiments sharing a common goal (family of experiments). In each case, particular synthesis techniques are applied (e.g., aggregated or individual patient meta-analysis), followed by the interpretation of results. During the interpretation, the researchers can identify moderator variables and generalize pieces of knowledge. Such findings may lead to a reformulation of the synthesis' goal or the conduction of a new cycle of experimental research.

\subsection{Experimental research concept model}

The conceptual model that this research revealed is not monolithic; in turn, it is composed of three sub-models (more accurately, viewpoints) aligned with the prominent experimental roles (research manager - RM, experiment manager - EM, and senior experimenter - SrEx). The viewpoints are structured according to the activities described in the EPW above, e.g., experiment definition, design, execution, etc.

\subsubsection{Research manager viewpoint}

The RM viewpoint is available in this \href{https://drive.google.com/file/d/1n3CHZbyjNovlgIhyTdCtN51f2X9Z-lwx/view?usp=sharing}{\ul{link}}. The RM has the following perspective:
\begin{itemize}
\item In the experiment definition phase, the RM manages the "problem definition" and the "hypothesis definition." 
\item During the experiment design phase, the RM takes care of the "human resource management" that could "run," "replicate," or be interested in "replicating" the experiment. The RM receives information regarding the "Expected results," "Execution context," and the "new variables of interest."
\item While the experiment runs, the RM monitors events; however, the data acquisition phase is transparent to her.
\item During the data analysis phase, the RM learns the "analysis results" and contributes to the definition of the "findings," "publications," and "pieces of knowledge."
\end{itemize}

\subsubsection{Experiment manager viewpoint}

The EM viewpoint is available in this \href{https://drive.google.com/file/d/1n3CHZbyjNovlgIhyTdCtN51f2X9Z-lwx/view?usp=sharing}{\ul{link}}. The perspective is as follows:

\begin{itemize}
\item In the experiment definition phase, the EM actively interacts with all experiment entities, hand in hand with the Senior Experimenter (SrEx). In the "hypothesis definition," the EM is only an observer.
\item In the experiment design phase, the EM focuses on improving the "experiment guides" regarding "instruments," "objects," and the "experimental kit". Additionally, the EM can analyze the context in which the experiment takes place.
\item During the experiment execution and data acquisition phases, the EM does not lose sight of the "events" that arise and could affect the experiment results. The EM also analyzes the "correctness" of the "collected items" and the "raw data."
\item In the data analysis phase, the EM cooperates with the "re-analysis" in search of new "variables of interest" to the RGUS. Additionally, the EM can observe the analysis results to suggest findings that could be useful in the publications.
\end{itemize}

\subsubsection{Senior experimenter viewpoint}

The SrEx viewpoint is available in this \href{https://drive.google.com/file/d/1RRDbJaav0XD7Oq9oG77VNTttvIZm89NL/view?usp=sharing}{\ul{link}}. The perspective is as follows:

\begin{itemize}
\item In the experiment definition phase, the SrEx observes the "experiment definition," hypothesis's "levels," and "metrics" in the first phase of the experimental investigation. The SrEx participates in the definition of the "hypothesis," "factors," and the "response variable."
\item In the experimental design phase, the SrEx interacts with a large number of entities; hence, it was necessary to divide such a phase into three sub-phases for this role: \textit{experimental operation design}, \textit{experiment design}, and \textit{artifacts generation}.
\begin{itemize}
\item During the \textit{experimental operation design}, the SrEx defines the "experimental objects" for each "experimental task," considering the minute details of each experimental object's characteristics. 
\item During the \textit{experiment design} proper, the SrEx establishes the execution context's "parameters," the "design type" with its "groups" and assigned "experimental subjects," the "levels" defined and their combination per "session" and "period," "experimental instruments," and the "measurement procedure." 
\item Finally, in the \textit{artifact generation}, the SrEx fundamentally develops the "run-kit," which includes the construction of the necessary instances of the "guides," "experimental instruments," and "experimental objects." Such "instances" mean tools for the "subjects" can execute the "experiment."
\end{itemize}
\item While the experiment runs, the SrEx records all the events produced to analyze their impact on the run's results.
\item During data acquisition, the SrEx records the resulting collected items, creates the raw data, and obtains measurements.
\item Finally, in the data analysis phase, the SrEx performs the analysis, obtains the findings, and prepares the publications with the RM and EM's help.
\end{itemize}

\subsection{Experimental process model}

\begin{figure*}[htbp]
\begin{center}
\includegraphics[trim=0 0 0 48,clip,width=0.95\textwidth]{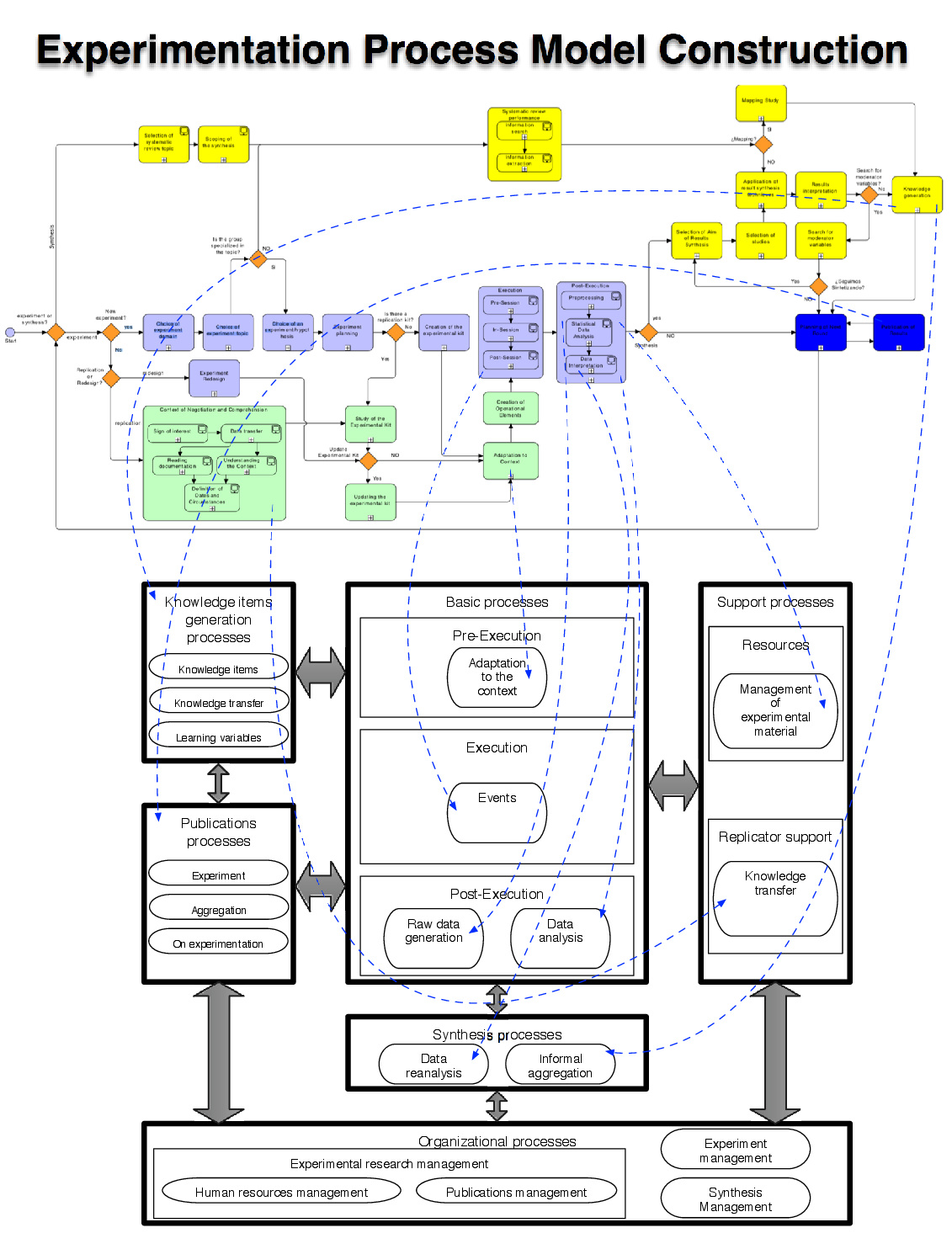}
\caption{Relationships between the \textit{experimental process workflow} (top) and the \textit{experimental process model} (down).}
\label{fig-EPM-construction}
\end{center}
\end{figure*}

The workflow and the conceptual models represent the experimental research process carried out in the RGUS. Other research groups may work differently. We have created an initial (generic, i.e., generally applicable) experimental process model (EPM) with the help of the most experienced experimenter at RGUS. The model is inspired in the software life cycle \cite{ISO-IEC-IEEE-12207}. We have adapted the concepts used in the ISO/IEC 12207 to the reality of SE experimentation. Most of the items have been taken from the EPW (see Figure \ref{fig-EPM-construction}); a lesser portion derives from the conceptual models. The complete process model can be obtained in this  \href{https://drive.google.com/file/d/1T6Nt4xKl5JdqdhFSRTeRdffFntLTJ_RB/view?usp=sharing}{\ul{link}}.

The EPM contains six groups of processes that define the SE experimental research life cycle: (1) Basic processes, (2) Support processes, (3) Knowledge items generation processes, (4) Publications processes, (5) Synthesis processes, and (6) Organizational processes. 

\subsubsection{Basic processes}

The basic processes involve fundamental activities that experimenters perform as RM, SrEx, and EM roles before, during, and after running an experiment. Before the experiment, the activities carried out are mainly focused on its design, e.g., adapting the experiment to the research context. During the experiment, experimenters pay attention to, e.g., the events that occur before, during, and after the experiment's execution session. The activities performed after the execution focus on the data, e.g., generating raw data and data analysis.

\subsubsection{Support processes}

The support processes include activities that the EM performs to manage the materials and support those interested in performing replications. The materials management's activities depend on the experiment's execution stage:
\begin{itemize}
\item The management of materials before the experiment's execution (e.g., experiment design document), 
\item the materials used during the experiment (e.g., templates for data collection), and 
\item the materials obtained at the end of the experiment (e.g., filled templates, subjects' code, etc.). 
\end{itemize}

Replication-support activities focus on guiding those interested in replicating the experiment.

\subsubsection{Knowledge items generation processes}

The knowledge items' generation processes consolidate activities carried out by the RM to manage the knowledge acquired after each experiment, e.g., the knowledge generated in each experimentation phase, the knowledge transferred to professionals, or the knowledge generated during the identification of moderator variables.
   
\subsubsection{Publications processes}

This group of activities focuses on reporting the experiment findings  (SrEx), aggregation of results (EM), and experimentation goals (RM).

\subsubsection{Synthesis processes}

The EM carries out the synthesis processes. There are different types of synthesis with different evidential levels, e.g., meta-analysis, re-analysis, and informal aggregations.

\subsubsection{Organizational processes}

The organizational processes include RM's management activities over the complete SE experimental research cycle, e.g.,  human resources and publications' management. Likewise, the EM also performs management activities on the experiment's materials. Finally, there are activities related to synthesis management that have not been assigned to any role yet.

\section{Discussion and related work}\label{sec-related}

This research aimed to find out how experimenters work in practice. We have obtained a set of models (process, workflow, conceptual) after conducting ethnographic research. Of course, these models have a strong resemblance to the~\textquotedblleft big picture\textquotedblright~that empirical researchers have about their research area. At a high level, the models (particularly the conceptual model) do not introduce innovations. However, a closer examination reveals exciting observations:
\begin{itemize}
\item \textbf{The experimental process is broader} than we usually recognize as~\textquotedblleft experimentation\textquotedblright. An experiment can be characterized as a small (or not that small) research project. It includes activities related to (see this \href{https://drive.google.com/file/d/1T6Nt4xKl5JdqdhFSRTeRdffFntLTJ_RB/view?usp=sharing}{\ul{link}}):
 \begin{itemize}
  \item management (resources, research lines, etc.),
  \item negotiation (e.g., with companies), 
  \item document/artifact support, 
  \item publication, etc.
 \end{itemize}
  
\vspace{3mm}

Although all empirical researchers are aware of these activities, it is surprising that they are seldom mentioned (if ever) in ESE literature. To the best of our knowledge, reference texts, e.g., \cite{Wohlin-2000-Experimentation-SE,Juristo-2001-SE-experimentation} \textbf{do not address these activities beyond marginal notes}. Such richness in the number and diversity of activities raises obvious consequences in experimental education and training.

\item One consequence of such diversity was revealed by the ethnographic study. It is possible, albeit unlikely, that one single experimenter performs all activities. These tasks are not inherently complex beyond the experimenters' abilities, but they demand time. In a project (research or not), task diversity is usually associated with the existence of roles with well-defined responsibilities. \textbf{Likewise in experimentation}. We identified three roles: (1) \textit{Research Manager}, (2) \textit{Experiment Manager} and (3) \textit{Senior Experimenter}. These roles probably have some connection to the RGUS' size; smaller (or larger) groups may have more generic/specialized roles. 

The concept of~\textquotedblleft role\textquotedblright~applied to experimentation is unusual but not completely new. Mohamed et al. \cite{Mohamed-1993-roles-ESE} proposed a framework to conduct ESE experiments. This framework was based on a merge between statistical and SE process concepts, \textbf{including explicit roles}.

\item In the same vein, the high-level conceptual model (see this \href{https://drive.google.com/file/d/1rGhYwH5I99ovmFGlX4FcLDCvd-y-ajp_/view?usp=sharing}{\ul{link}}) could be endorsed by any experimenter (with some complaints depending on his/her area of specialization, as it happened during the ethnography. In this regard, see below.

When the concept of role appears in the scene, the high-level model is just \textbf{an abstraction of the concrete tasks performed during an experimental instance}. The high-level model evokes the viewpoint of the \textit{senior experimenter} (see this \href{https://drive.google.com/file/d/1RRDbJaav0XD7Oq9oG77VNTttvIZm89NL/view?usp=sharing}{\ul{link}}), but with fewer details. In turn, the \textit{experimental manager} (see this \href{https://drive.google.com/file/d/1Du5viBUKfrNehlLs9U0WeknKAQzstxHe/view?usp=sharing}{\ul{link}}) and the \textit{research manager} (see this \href{https://drive.google.com/file/d/1n3CHZbyjNovlgIhyTdCtN51f2X9Z-lwx/view?usp=sharing}{\ul{link}}) pay little attention to the design and execution details because the \textit{senior experimenter} takes care of that. 

Likewise, the \textit{experiment manager} is not affected by the experimental research in its global context, e.g., collaboration with other groups and replication. Such a concern fits the \textit{research manager}'s role. 

\item The process, workflow, and conceptual models were built to offer a coherent picture purposedly. However, as we indicated in section \ref{subsubsec-focus-groups}, we were unable to manage the existence of different families of experiments. When experimenters working in different families get together, the models automatically diverge. 

\item  The process, workflow, and conceptual models represent the maximum achievable consensus, obtained during the role-oriented meetings. However, we are conscious that \textbf{further research would be necessary at the family level, to find out the peculiarities of each research area}. Such an inquiry could have some impact. It is likely that, e.g., replications would benefit from the specialized, fine-grained knowledge present at the family level. The abstraction of such specialized knowledge into coarse-grained models (like \href{https://drive.google.com/file/d/1T6Nt4xKl5JdqdhFSRTeRdffFntLTJ_RB/view?usp=sharing}{\ul{this}} model) hides the details and makes replication harder. Other authors have already pointed out that the replication problem is a knowledge-transfer problem \cite{Shull-2004-Knowledge-sharing-issues-SE}.

\end{itemize}

The models, and the ethnographic experience, have generated some differential characteristics more, e.g., context adaptation, iterative improvement, etc. Some of these characteristics have been pointed out in previous research, e.g., \cite{Mohamed-1993-roles-ESE,Sjoberg-2005-survey-experiments-SE,Shull-2004-Knowledge-sharing-issues-SE}. 

We do not wish to provide a detailed account of these differential characteristics herein; that would require a publication of a different character. We aimed to observe how experimenters work, and we found out that they perform other tasks than the ones portrayed in textbooks. This is our key contribution. 
\section{Threats to Validity}\label{sec-threats}

The study's validity is the degree to which the research has been conducted with transparency and the absence of research biases. We have followed Runeson et al.'s recommendations \cite[p. 71–73]{Runenson-2012-case-study-SE} to design the ethnographic research and describe the threats to validity. Runeson et al., following Yin \cite{Yin-2009-case-study}, classify threats in four groups: (1) construct validity, (2) internal validity, (3) external validity, and (4) reliability.

\subsection{Construct validity}

This category deals with the degree of agreement between the constructs developed for the study (concerning the research questions), and what the researcher has in mind. For example, if interview questions are not interpreted in the same way by the researcher and the experimenters, there is a threat to construct validity. 

The ethnographic method is particularly suitable to fight against this type of threat. During the research, we used several information-acquisition techniques to create constructs, e.g., reading, observation, interviews, etc. The different approaches increase the chances to identify/solve misunderstandings between the researchers and the RGUS. We also prevented construct validity threats using reliable information sources, e.g., standard literature and the RGUS' experimental material. 

Finally, the constructs were validated iteratively by experimenters and researchers during the research project lifespan.

\subsection{Internal validity}

Internal validity addresses the credibility of the causal relationships found during the research. This threat does not apply this ethnographic study, as we do not aim to identify causal relationships but, at most, correlations between phenomena, e.g., when we claim that the number and diversity of experimental tasks explain the existence of roles. Furthermore, most of the findings that we provide are of descriptive nature, e.g., the disagreements among experimenters of different experimental families.

There is a second threat to internal validity (it can be seen as the foundation of causality) related to inferences \cite{Yin-2009-case-study}. Carrying out a qualitative study implies making inferences (inductive and abductive) that involve phenomena that cannot be directly observed (the constructs). This threat does influence an ethnographic study since the researcher becomes an active entity in the environment under study. If the researcher makes an incorrect interpretation of facts or phenomena, the internal validity could be threatened. We mitigated this threat validating the researchers' observations, claims, and conclusions with the RGUS' experimenters.

\subsection{External Validity}

External validity represents the degree of certainty of the findings obtained in an investigation being generalizable, i.e., applied to different contexts where they have been obtained. This threat is prominent in this study since it was carried out in a single instance (a research group) of the population under study (SE research groups that apply empirical methods). As a countermeasure, we chose a highly representative ESE research group.

\subsection{Reliability}

The reliability of a study is the ease with which the research activities, e.g., the data collection procedure, and the results obtained, e.g., the existence of roles, can be reproduced by other studies that apply the same methodology in the same RGUS; or more likely, in similar ESE research groups.

We have faithfully followed the research methodology outlined in Section~\ref{sec-research-method} to ensure the reliability of this research. Moreover, we have disclosed all intermediate and final results in a repository, which is available under request, to guarantee that the research has been conducted with transparency.

\section{Conclusions}\label{sec-conclusions}

ESE researchers know that experimentation, as described in textbooks and guidelines, is only an approximation to the daily work in labs and the field. This study reveals some deviations found in a concrete research group. Such deviations do not seem arbitrary but necessary from a systemic viewpoint, e.g., different people have different abilities and perform different tasks in a project; consequently role-specialization shall appear \textit{naturally} in ESE.

We do not claim that our findings are commonplace in ESE. They may be highly specific to the research group in which we performed the ethnographic study. However, some clues evidence that peculiar behaviors are usual in labs and, probably the field as well. 

For a now long time, natural sciences researchers complain about the replication problem \cite{hines2014sorting}. At least in the sciences (maybe not in ESE), replication frequently fails due to changes in the experimental setup and conduction. These changes are motivated by the researchers' tacit knowledge \cite{Polanyi-1996-tacit-k,Shull-2002-replicating-SE-experiments-tacit-k}, and consequently such changes are not reported in protocols or experimental reports. Variations in experiments happen so frequently that in some disciplines a (comic) term has been coined: \textquotedblleft lab mythology\textquotedblright~\cite{ruben2011experimental,loukides2015beyond}. Our study has shown that such tacit knowledge is present in ESE, and it manifests at a rather superficial level, e.g., when researchers from different experimental families discuss their mental models.

Furthermore, anecdotic evidence suggests, e.g., conversations with colleagues, suggest that this ''lab mythology'' can be somewhat present in SE as well. We all know (and perhaps done as well) that experiments are not really conducted exactly the same as they are reported. It does not mean that we cheat; in turn, we know from experience that some practices work and we apply them regardless they are mainstream or not.

We believe that the~\textquotedblleft lab mythology\textquotedblright~is worth studying. If the differential characteristics observed are research-group specific, the inquiry would have come to an end. However, it is also possible that widespread patterns of behavior show up. Bringing such knowledge to the foreground could contribute to the improvement of experimental practice.






\bibliographystyle{elsarticle-num}
\bibliography{bibliography}







\end{document}